\documentclass[10pt, conference, letterpaper]{IEEEtran}
\IEEEoverridecommandlockouts
\usepackage{balance}
\usepackage{amssymb}
\usepackage{stmaryrd}
\usepackage{color}
\usepackage{multirow}
\usepackage{graphicx,times}
\usepackage{epstopdf}
\usepackage{indentfirst}
\usepackage{CJK}
\usepackage{amsmath}
\usepackage{amsfonts}
\usepackage{txfonts}
\usepackage{mathrsfs}
\usepackage{subfigure}
\usepackage{graphicx}
\usepackage{theorem}
\usepackage{url}
\usepackage{microtype}
\usepackage{fancyhdr}  

\def\BibTeX{{\rm B\kern-.05em{\sc i\kern-.025em b}\kern-.08em
		T\kern-.1667em\lower.7ex\hbox{E}\kern-.125emX}}

\begin{document}
	
\title{Lightweight Sybil-Resilient Multi-Robot Networks by Multipath Manipulation}
\author{\IEEEauthorblockN{Yong Huang{$^{^\dagger}$}, Wei Wang{${^\ast}{^{^\dagger}}$}, Yiyuan Wang{$^{^\dagger}$}, Tao Jiang{$^{^\dagger}$}, Qian Zhang{${^\S}$}}\IEEEauthorblockA{{$^{^\dagger}$}School of Electronic Information and Communications, Huazhong University of Science and Technology\\ {${^\S}$} Department of Computer Science and Engineering, Hong Kong University of Science and Technology\\Email: \{yonghuang, weiwangw, yiyuanwang, taojiang\}@hust.edu.cn, qianzh@cse.ust.hk}
	\thanks{${^\ast}$The corresponding author is Wei Wang (weiwangw@hust.edu.cn).}
}	

\maketitle
\thispagestyle{fancy}          
\fancyhead{}                     
\chead{IEEE INFOCOM 2020 - IEEE International Conference on Computer Communications}
\cfoot{} 
\renewcommand{\headrulewidth}{0pt}     
\renewcommand{\footrulewidth}{0pt}

\begin{abstract}
Wireless networking opens up many opportunities to facilitate miniaturized robots in collaborative tasks, while the openness of wireless medium exposes robots to the threats of Sybil attackers, who can break the fundamental trust assumption in robotic collaboration by forging a large number of fictitious robots. Recent advances advocate the adoption of bulky multi-antenna systems to passively obtain fine-grained physical layer signatures, rendering them unaffordable to miniaturized robots. To overcome this conundrum, this paper presents ScatterID, a lightweight system that attaches featherlight and batteryless backscatter tags to single-antenna robots to defend against Sybil attacks. Instead of passively ``observing'' signatures, ScatterID actively ``manipulates'' multipath propagation by using backscatter tags to intentionally create rich multipath features obtainable to a single-antenna robot. These features are used to construct a distinct profile to detect the real signal source, even when the attacker is mobile and power-scaling. We implement ScatterID on the iRobot Create platform and evaluate it in typical indoor and outdoor environments. The experimental results show that our system achieves a high AUROC of 0.988 and an overall accuracy of 96.4\% for identity verification.
\end{abstract}


\section{Introduction}

The development of wireless technologies has promised to facilitate effective collaboration among a team of small and agile robots, which enables a broad range of intriguing applications, such as surveillance~\cite{rybski2002performance}, consensus~\cite{jadbabaie2003coordination}, and search and rescue~\cite{zhang2006novel}. Although the openness of wireless medium delivers on the promise for efficient and agile collaboration, it also exposes robots to cyber attacks. A particular detrimental attack in multi-robot networks is the Sybil attack, which easily breaks the fundamental trust assumption in robotic collaboration by forging a bunch of fake identities to gain a disproportionate influence in the network~\cite{newsome2004sybil}. 


However, due to the ad hoc, dynamic, and miniaturized characteristics of robotic platforms, the Sybil attack mitigation in multi-robot networks still remains to be a challenging issue. Traditional pre-shared key (PSK) management schemes presume prior trust among network nodes~\cite{chan2003random,ramkumar2005an}, which are difficult to implement in ad hoc robotic networks where robots often go in and out. Alternatively, research efforts~\cite{demirbas2006an,xiao2009channel-based,liu2015the,faria2006detecting,xiong2013securearray,gil2017guaranteeing} measure received signals strength indicator (RSSI), channel state information (CSI) and angle of arrival (AoA) features in wireless physical layer (PHY) to verify the spatial uniqueness of each node. However, RSSI and CSI based techniques~\cite{demirbas2006an,xiao2009channel-based,liu2015the,faria2006detecting} not only need collaboration among multiple receivers or antennas, but also require all nodes to be stationary or semi-stationary. Although fine-grained AoA signatures can be extracted to detect nodes in close proximity under dynamic channels~\cite{xiong2013securearray,gil2017guaranteeing}, these approaches either rely on large multi-antenna arrays~\cite{xiong2013securearray} or require unnecessary robotic motions, such as in-place spin with two antennas~\cite{gil2017guaranteeing}. They are ill-suited to a team of robots individually with limited payload and hardware capabilities. 

\begin{figure}
	\centering
	\includegraphics[width=0.88\linewidth]{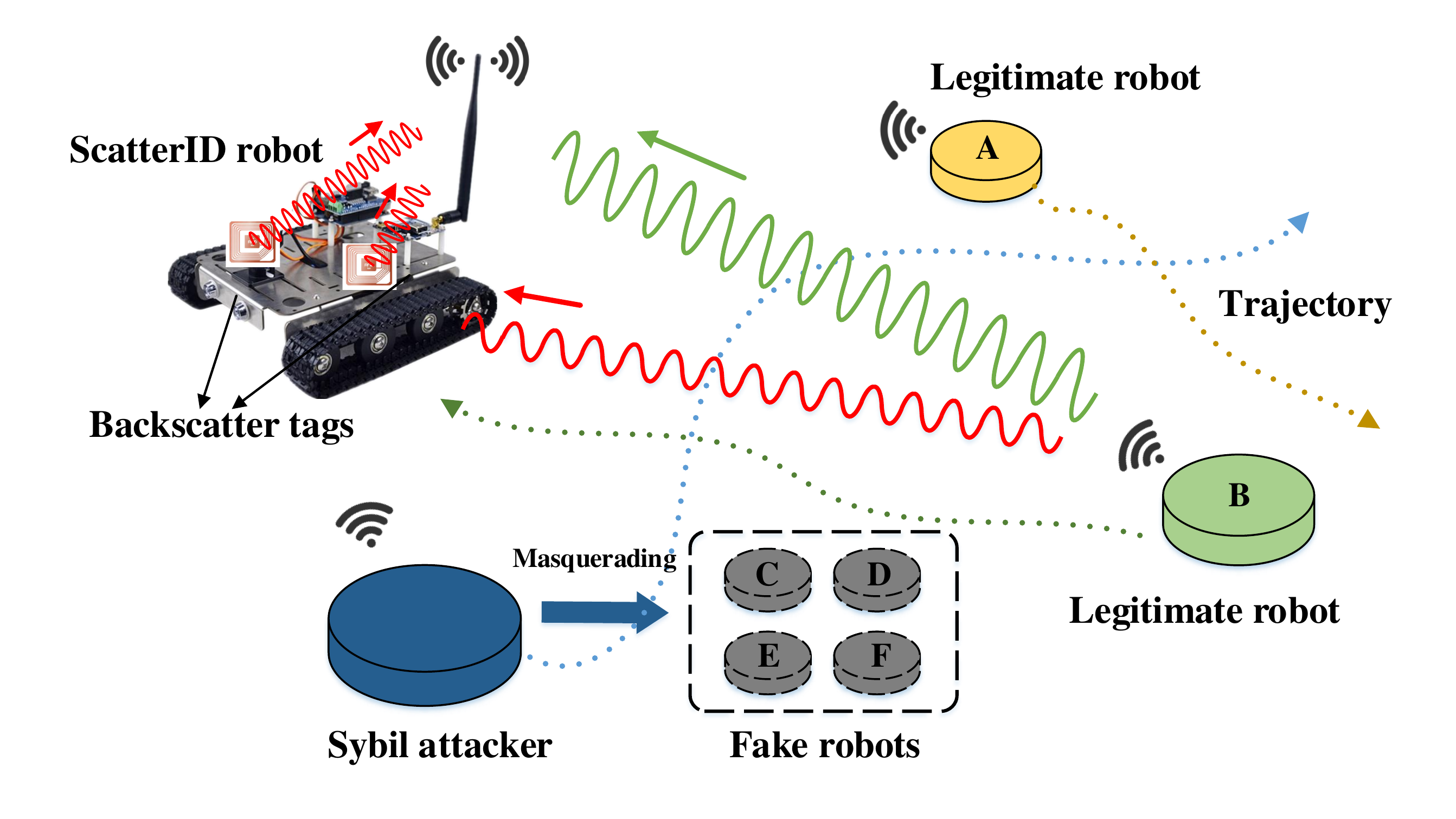}
	\caption{Illustration of Sybil attacks in a multi-robot network.}
	\label{fig:observation}
\end{figure}

In this paper, we argue that the fundamental hurdle in realizing lightweight Sybil-resilient solutions lies in that these RF-based innovations focus on passively ``observing'' signal propagation signatures, which require bulky multi-antenna systems to capture fine-grained information. This paper explores a new approach: can we instead actively ``manipulate'' multipath propagation, to make conventionally multi-antenna-exclusive signatures also obtainable to a single antenna? If we could alter multipath propagation by just attaching several featherlight and batteryless backscatter tags to existing single-antenna robots, it would not require any hardware modification or incur load burden to the robots. 

Toward this end, we present ScatterID, a lightweight system that attaches backscatter tags to single-antenna robots to defend against Sybil attacks. Our fundamental insight is that when backscatter tags communicate by intermittently absorbing and reflecting ambient signals, the multipath between a pair of transceivers changes correspondingly~\cite{liu2013ambient}. Such fast changes, i.e., reflections from tags, provide unique spatial properties of the communication pair. In particular, backscattered signal strengths are highly correlated to distances between transceivers with respect to tags. By affixing several tags on a robot, as shown in Fig.~\ref{fig:observation}, other robots' trajectory information can be conveyed in backscattered signal traces. The spatial correlation indicates that similar backscatter signatures come from the same moving robot with a high probability. In this way, a ScatterID robot can leverage backscatter tags to extract unforgeable PHY IDs for discriminating between legitimate robots and Sybil attackers.

We realize the above idea by tackling the following two challenges.

\textit{1) How to exploit backscatter tags to construct sensitive signal profiles for hardware-constrained and dynamic robots?} With merely one antenna, a robotic platform cannot acquire fine-grained signal signatures, such as AoA, for reliable attack detection in dynamic channel states. To deal with this challenge, we take sequential backscatter signatures of each robotic transmitter as a signal profile to characterize its long-term spatial information. Specifically, when hearing data transmission from a surrounding robot, a ScatterID robot controls its tags to reflect wireless signals in turn. These backscattered signals are unique to the transmitter's location. In this way, the ScatterID robot continuously reflects a series of transmissions to obtain a signal profile that is sensitive to the transmitter's trajectory. 

\textit{2) How to effectively measure the similarity between signal profiles under power-scaling attacks?} In order to emulate different robots, a Sybil attacker can scale its transmission power with a varying coefficient for each identity, which consequently leads to different signal profiles. To defend against such an attack, we propose an effective algorithm for scaling-resilient similarity measurement. In particular, it first normalizes each signal profile to mitigate the scale coefficient. Then, it leverages an adjusted cosine metric to effectively measure the distance between two normalized profiles. Based on the measured distances, a logistic regression model is exploited to learn two similarity probabilities for a pair of robots. If both two probabilities approach one, the robotic pair are very likely forged by the same robot.

\textbf{Summary of Results.} We implement our system using iRobot Create robots, GNURadio/USRP B210 and backscatter tags using commercial off-the-shelf circuit components, and extensively evaluate our system in typical indoor and outdoor environments. The evaluation results show that our system achieves a high area under the receiver operating characteristic curve (AUROC) of 0.988 and an overall accuracy of 96.4\% for identity verification. Moreover, it can successfully detect 97.6\% of fake robots and meanwhile mistakenly reject just 5.1\% of legitimate robots.

\textbf{Contributions.} The main contributions of this work are summarized as follows. First, we show that featherlight and batteryless backscatter tags can be exploited to create fine-grained PHY features, which are obtainable to single-antenna robots and thus offer a compelling alternative to the ones using bulky or specialized multiple antennas. Second, we propose ScatterID, a lightweight system that is resilient to mobile Sybil attackers with power-scaling ability. Third, we implement ScatterID on commercial off-the-shelf robotic platforms and conduct extensive experiments in indoor and outdoor environments to demonstrate the effectiveness and robustness of our system.

The remainder of this paper is organized as follows.  Section~\ref{sec:motivation} introduces Sybil attacks in multi-robot networks and signatures from backscatter tags for Sybil attack detection. Section~\ref{sec:system design} elaborates on the system design. Implementation and evaluation are described in Section~\ref{sec:experiment}. Section~\ref{sec:related work} gives a literature review, followed by conclusion in Section~\ref{sec:conclusion}.

\section{Backscatter-Enabled Signatures for Sybil Attack Detection}\label{sec:motivation}
 
\subsection{Sybil Attacks in Multi-Robot Networks}
We focus on a general multi-robot network, where a team of miniature mobile robots coordinates their actions by exchanging information with each other in an ad hoc manner~\cite{jadbabaie2003coordination,cunningham2012fully}. In the robotic team, each robot has limited payload and hardware capabilities and is equipped with only one antenna for data transmission and reception. Moreover, all mobile robots have distinctive moving paths in a certain environment, since multi-robot systems are usually spatially distributed to complete given tasks in a time-efficient way~\cite{yan2013survey}.

In the robotic network, Sybil attacks occur when a selfish robot masquerades multiple fictitious robots by broadcasting messages with different identities. The malicious robot is considered as a Sybil attacker. With many fake identities, the attacker can obtain more attention from neighboring robots and further brings a disproportionate influence into the whole network. Additionally, the Sybil attacker can launch power-scaling attacks by varying its transmission power in purpose to fool other robots. The robots that are not spawned are considered as legitimate robots. 

\subsection{Characterizing Backscattered Signals} \label{sub:analysis}
Backscatter tags can provide fine-grained PHY signatures that contain spatial properties of a transmitter-receiver pair. In particular, backscatter tags enable wireless communication by intermittently absorbing and reflecting ambient radio signals, which changes the multipath propagation between the transmitter and receiver. Such fast changes are highly correlated to the locations of transceivers with respect to tags. Since the multipath propagation between the transceiver pair is constant, we consider it as a compound path that is equivalent to a direct path in free space. Hence, we can focus on the multipath changes incurred by backscatter tags. According to Friis path loss theory~\cite{rao2006theory}, the reflected signal strength $ P_{reflected} $ by a backscatter tag can be expressed as
\begin{align} \label{friis path loss}
	P_{reflected}=\frac{P_t G_t}{4 \pi d_{t}^{2}} \times \frac{\lambda^{2} G_r}{16 \pi^2 d_{r}^{2}} \times T(\lambda,G_{tag},\Gamma),
\end{align}
where $ P_t $ is the transmission power, $ d_t $ and $ d_r $ the tag's equivalent distances to the transmitting and receiving antennas computed based on the compound direct path, $ G_t$ and $ G_r $ the transmitting and receiving antenna gains, respectively. Moreover, $ T(\cdot) $ is a function of the wavelength $ \lambda $, the antenna gain $ G_{tag} $ of the tag and its reflection coefficient $ \Gamma $, which indicates the amplitude ratio of the incoming signal and the reflected signal. Based on Eq.~\eqref{friis path loss}, we can find that the backscattered signal strength $ P_{reflected} $ is highly dependent on the transmission power and the relative distances of the tag in terms of transmitting and receiving antennas.

By affixing a backscatter tag on a mobile robot, $ d_r $ is a fixed distance, and $ P_t $ and $ d_t $ are only variables in the Eq.~\eqref{friis path loss}. Therefore, the signal strength reflected by the tag highly correlates with transmission power and the distance between the robotic communication pair. With a constant transmission power $ P_t $, we can construct a spatial-related signature relying on reflected signals from multiple backscatter tags.

\subsection{Feasibility Study}

\begin{figure}
	\centering
	\includegraphics[width=0.88\linewidth]{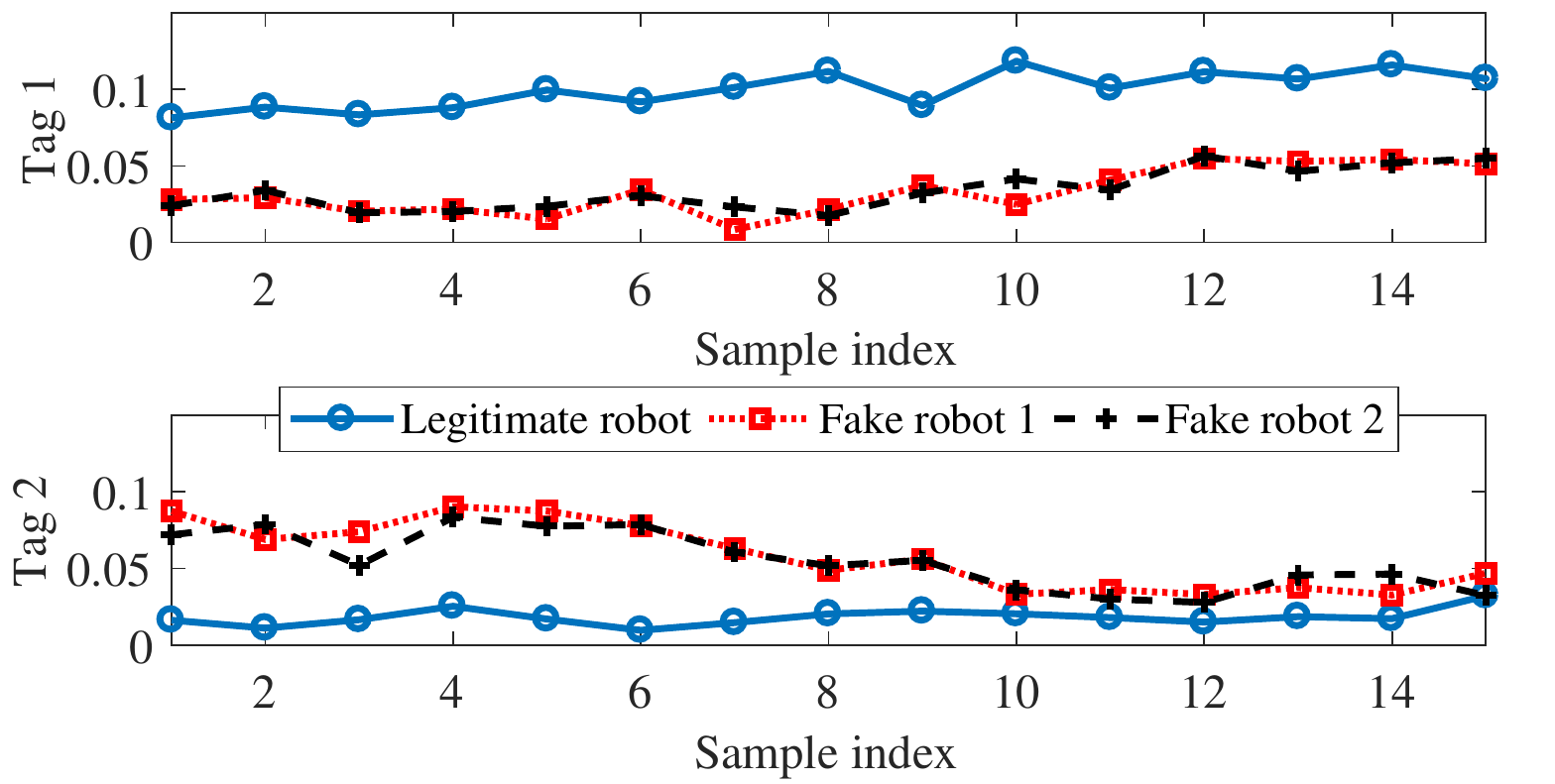}
	\caption{Backscattered signal traces reflected by two tags from one legitimate and two fake robots.}
	\label{fig:traces}
\end{figure}

To gain a better intuition, we perform a set of preliminary experiments to verify the above idea using backscatter tags, moving robotic platforms and USRP nodes. To avoid interference between the carrier and reflected signals, we implement frequency-shift backscatter tags based on recent work~\cite{kellogg2015wi-fi,zhang2016hitchhike,wang2017fm}. In the experiment, we attach two tags to one USRP node who logs backscattered signal traces in a fixed position. We place two USRP nodes on different moving platforms to emulate one legitimate robot and one Sybil attacker with two different identities, respectively. Both the legitimate robot and Sybil attacker transmit packets with the same transmission power while moving around in different trajectories. We extract the backscattered signal traces by two tags and plot the results in Fig.~\ref{fig:traces}. 

From Fig.~\ref{fig:traces}, we observe that for two fake robots, their backscattered signal traces are nearly identical to each other. This is because that from the same moving robot, their signals experience similar propagation paths that are created by two tags. Whereas, for a pair of legitimate and fake robots, their signal propagation to backscatter tags are different due to distinctive trajectories. Thus, their backscattered signal traces are uncorrelated with each other. This observation verifies that the reflected signals from backscatter tags are highly sensitive to robotic trajectories, and can be further utilized to construct a unique signal profile for each moving robot. 

\begin{figure}
	\centering
	\includegraphics[width=0.88\linewidth]{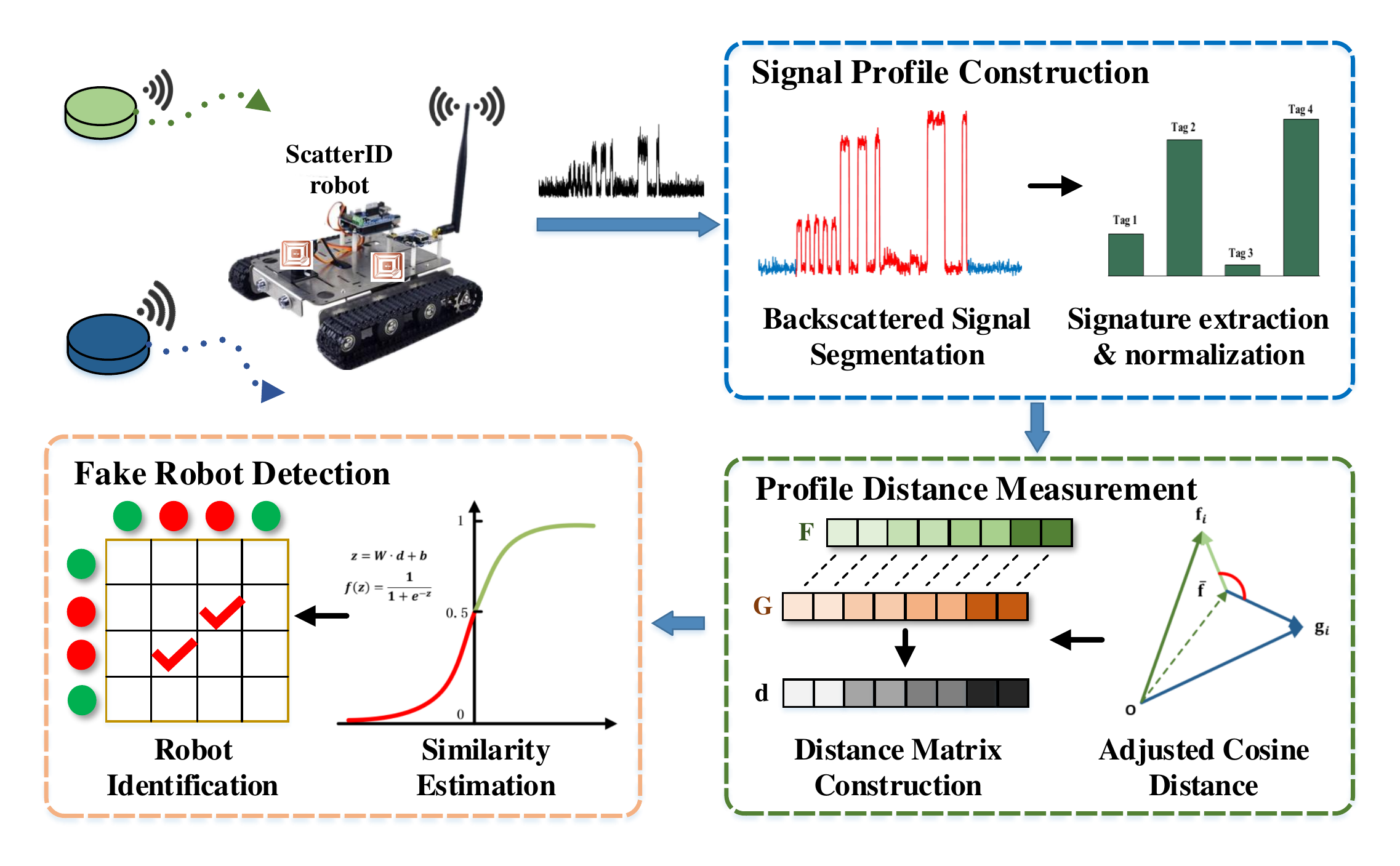}
	\caption{System architecture. It contains three components: Signal Profile Construction, Profile Distance Measurement and Fake Robot Detection.} 
	\label{fig:system-overview}
\end{figure}

\section{System Design}\label{sec:system design}

\subsection{System Overview} 
ScatterID is a lightweight system that provides effective resilience to Sybil attacks in multi-robot networks. To defend against Sybil attackers, ScatterID attaches featherweight and batteryless backscatter tags to a single-antenna robot to reflect data packets from its neighboring robots and creates unforgeable PHY IDs for them based on sensitive reflected signatures. Hence, our system does not require any hardware modification or incur much load burden to miniaturized robots.

As illustrated in Fig.~\ref{fig:system-overview}, ScatterID takes as input consecutive backscattered signals of other robots and outputs their identity legitimacy. The core of our system includes three components -- \textit{Signal Profile Construction}, \textit{Profile Distance Measurement} and \textit{Fake Robot Detection}.
\begin{itemize}
	\item  \textbf{Signal Profile Construction.} This component first segments the backscattered signal from the received signal, and then effectively extracts the reflected signals by all tags as a multipath signature. Then, a signal profile can be obtained as a sequence of normalized signatures for each robot.
	\item  \textbf{Profile Distance Measurement.} After profiling neighboring robots, this component measures the distance between any two signal profiles with an adjusted cosine metric and then outputs a distance matrix for subsequent similarity measurements.
	\item  \textbf{Fake Robot Detection.} This component leverages the logistic regression model to estimate the similarities between any two robots based on their measured distances. According to a matrix of similarities, the legitimacy of all robots can be identified.
\end{itemize}

\subsection{Signal Profile Construction}
The first step of ScatterID is to construct sensitive signal profiles for other robots. These profiles are carefully constructed based on a sequence of backscattered signals to provide unique spatial properties of moving robots. 

\begin{figure}
	\centering
	\includegraphics[width=0.88\linewidth]{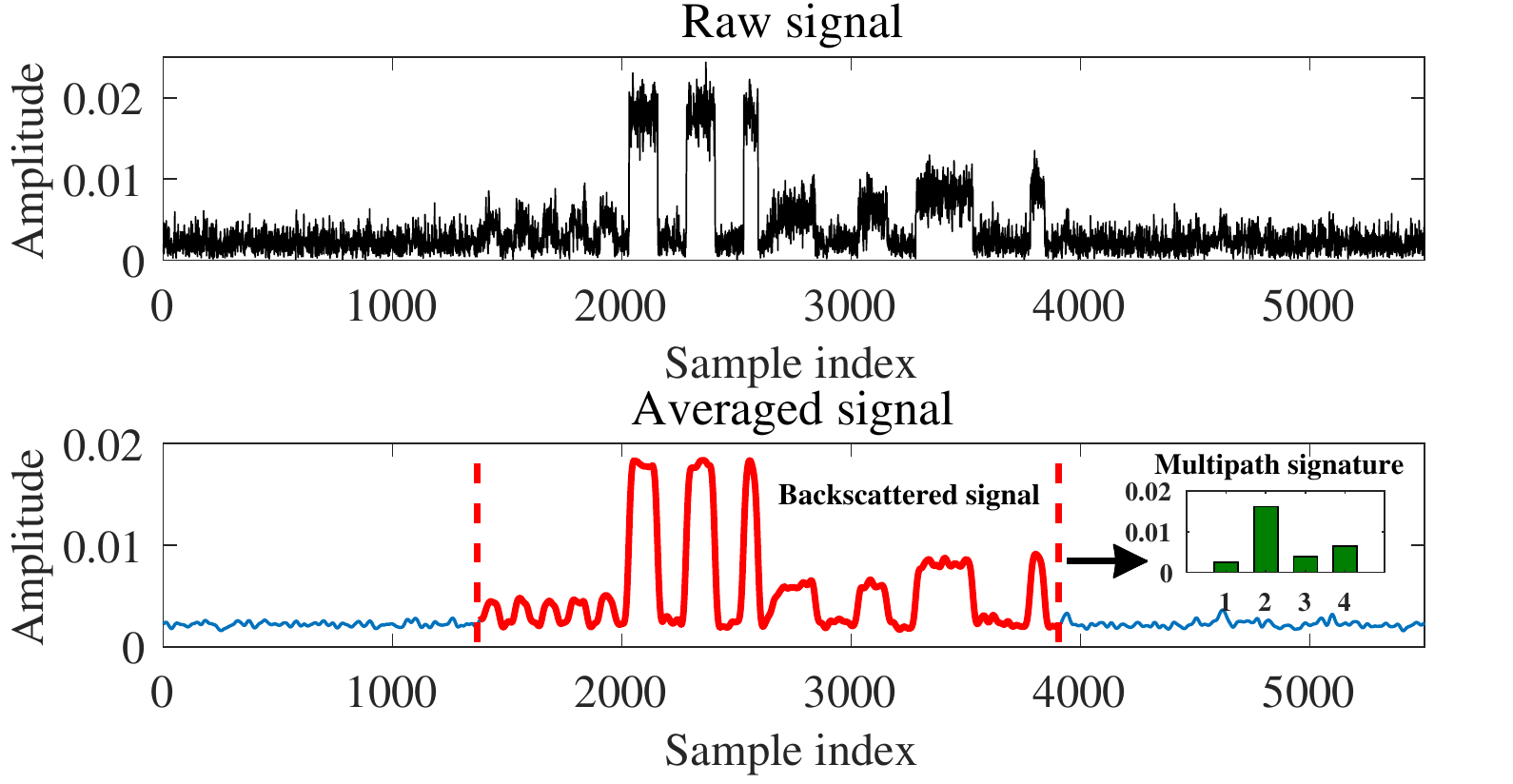}
	\caption{Backscattered signal segmentation and signature extraction. The red segment represents a backscattered signal, and the green bars indicate extracted backscatter signatures.}
	\label{fig:backscatteredsignalextraction}
\end{figure}

\textbf{Backscattered Signal Segmentation.} To update backscattered signals, a ScatterID robot controls all tags to reflect data packets sent from each robot and thereafter extracts a segment of backscattered signal from received signal traces periodically. Since the received signal may contain the component without tags' reflections, it is highly desirable to detect and segment the backscattered component from the received signal. 

To achieve this goal, our system first adopts a moving average method~\cite{liu2013ambient} to smooth the received signal. Fig.~\ref{fig:backscatteredsignalextraction} plots an instance of a received signal and the corresponding smoothing result. As the figure shows, such smoothing operations enable our system to effectively remove signal noise and reliably decode transmitted message from tags. Next, we leverage tags' transmitted message to determine the start and end of backscattering in the averaged signal. Specifically, the transmitted message is a known binary array of zeros and ones, which are encoded in the backscattered signal to differentiate each tag's reflection from others. Thereby, the backscattered component of the averaged signal and the tags' binary array are highly correlated. Hence, we can correlate the averaged signal $ \mathbf{s}(n) $ with the binary array $ \mathbf{i}(n) $ to detect when tags begin reflecting signals. In particular, the correlation result $ \mathbf{c}(n) $ can be written as 
\begin{align}\label{coorelation}
	\mathbf{c}(n)=\sum^M_{m=1} \mathbf{s}(n+m-1) \times \mathbf{i}(m),
\end{align}
where $ M $ is the length of $ \mathbf{i}(n) $. According to Eq.~\eqref{coorelation}, $ \mathbf{c}(n) $ has the highest peak when the backscattered signal and $ \mathbf{i}(n) $ completely overlap. Thus, our system identifies the backscattering start $ t_{start} $ by finding $ n$ with the maximum value in $ \mathbf{c}(n) $. Moreover, as the length of the backscattered signal equals to $ M $, we can accordingly obtain the end of backscattering as $ t_{end} = t_{start} + M $. With $ t_{start} $ and $ t_{end} $, our system is able to accurately segment the backscattered signal $ \mathbf{B} $ as
\begin{align}
	\mathbf{B}=(\mathbf{b}_{1},\mathbf{b}_{2},\cdots, \mathbf{b}_{K}),
\end{align}
where $ \mathbf{b}_{i} $ is the reflected component by the $i^{\text{th}}$ tag and $ K $ is the number of tags used in our system.

\textbf{Signature Extraction.} Now that we have picked out the backscattered signal, we take the next step to extract a multipath signature from it. As depicted in Fig.~\ref{fig:backscatteredsignalextraction}, the backscattered signal consists of not only the tags' reflections but also the unwanted reflections from ambient environments. To extract reliable multipath features from each tag, ScatterID leverages the fact that the difference between reflected and non-reflected samples is caused by the tags' reflections. Therefore, by subtracting two kinds of samples, we can effectively eliminate environmental reflections and thereafter obtain the reflections from all tags. Mathematically, with the backscattered signal of the $i^{\text{th}}$ tag $ \mathbf{b}_{i}(n) $, we assume that there are $ N_1 $ reflected and $ N_0 $ non-reflected samples. The reflection of the $i^{\text{th}}$ tag $ p_i $ can be computed by
\begin{align}\label{power extraction}
	p_i = \frac{1}{N_1} \sum^{N_1}_{n=1} \mathbf{b}_{i}(n|ref.)-\frac{1}{N_0} \sum^{N_0}_{n=1} \mathbf{b}_{i}(n|non-ref.).
\end{align}

As analyzed in Section~\ref{sec:motivation}, the tag's reflection $ p_i $ highly depends on the distance between the tag and signal resource. Thus, with the assistance of multiple backscatter tags, our system can construct a multipath signature that provides location information of the signal resource. Formally, based on $ K $ tags' reflections, a multipath signature $\mathbf{p} $ can be expressed as
\begin{align}
	\mathbf{p}=(p_1, p_2, \cdots, p_K).
\end{align}
Considering that $ p_i \geq 0  $, $ \mathbf{p} $ is a nonnegative vector in the feature space.

\textbf{Signature Normalization.} After obtaining a multipath signature $ \mathbf{p} $, we proceed to normalize it to deal with power-scaling attacks, where a clever attacker changes its transmission power to emulate different fake robots. Specifically, according to Eq.~\eqref{friis path loss}, by intentionally varying $ P_t $ in each transmission, the attacker can manipulate $ \mathbf{p} $ with a different scale coefficient $ \alpha $ for each identity. Such an operation scales the reflected signals by all tags and further leads to a scaled signature $  \mathbf{p}' = (\alpha p_1, \alpha p_2, \cdots, \alpha p_K) $, which consequently renders our system ineffective in discovering two fake robots. 

To obtain an unaffected signature with respect to power-scaling attacks, we normalize each $ K $-dimensional vector $ \mathbf{p} $ via dividing it by its L2 norm $ \left\|\mathbf{p}\right\|_2 $ and acquire a normalized multipath signature $ \mathbf{f} $ as
\begin{align} \label{normalized signature}
	\mathbf{f}  =\frac{\mathbf{p}}{\left\|\mathbf{p}\right\|_2}= \frac{ ( p_1,  p_2, \cdots, p_K)}{ \sqrt{p_1^2 + p_2^2 + \cdots + p_K^2}}
	 = (f_1, f_2, \cdots, f_K),
\end{align}
where $ f_i \geq 0 $ is the normalized reflection of the $i^{\text{th}}$ tag. This transformation can help our system eliminate the scale coefficient $ \alpha $ in the scaled signature $ \mathbf{p}' $ and therefore makes any multipath signature resilient to power-scaling attacks.

\textbf{Profile Construction.} To this end, ScatterID can construct a signal profile for each robot. Due to the mobility of both ScatterID and neighboring robots, multipath signatures from the same Sybil attacker may have a poor similarity. To deal with this issue, our system periodically extracts a multipath signature from each robot's backscattered signal traces and takes a sequence of successive signatures as a signal profile for capturing its trajectory information. For fake robots forged by the same moving attacker, their backscattered signal traces have the same trend and are nearly identical to each other, as depicted in Fig.~\ref{fig:traces}. Hence, by characterizing long-term information, their signal profiles will be similar to each other. Formally, a signal profile $ \mathbf{F} $ of each moving robot can be expressed as 
\begin{align}
\mathbf{F}=[\mathbf{f}_1; \mathbf{f}_2; \cdots; \mathbf{f}_L],
\end{align}
where $ \mathbf{f}_i $ denotes the $i^{\text{th}}$ normalized signature and $ L $ is the number of signatures.

\subsection{Profile Distance Measurement} 

\begin{figure}
	\centering
	\subfigure[Cosine distance.]{
		\includegraphics[width=0.36\linewidth]{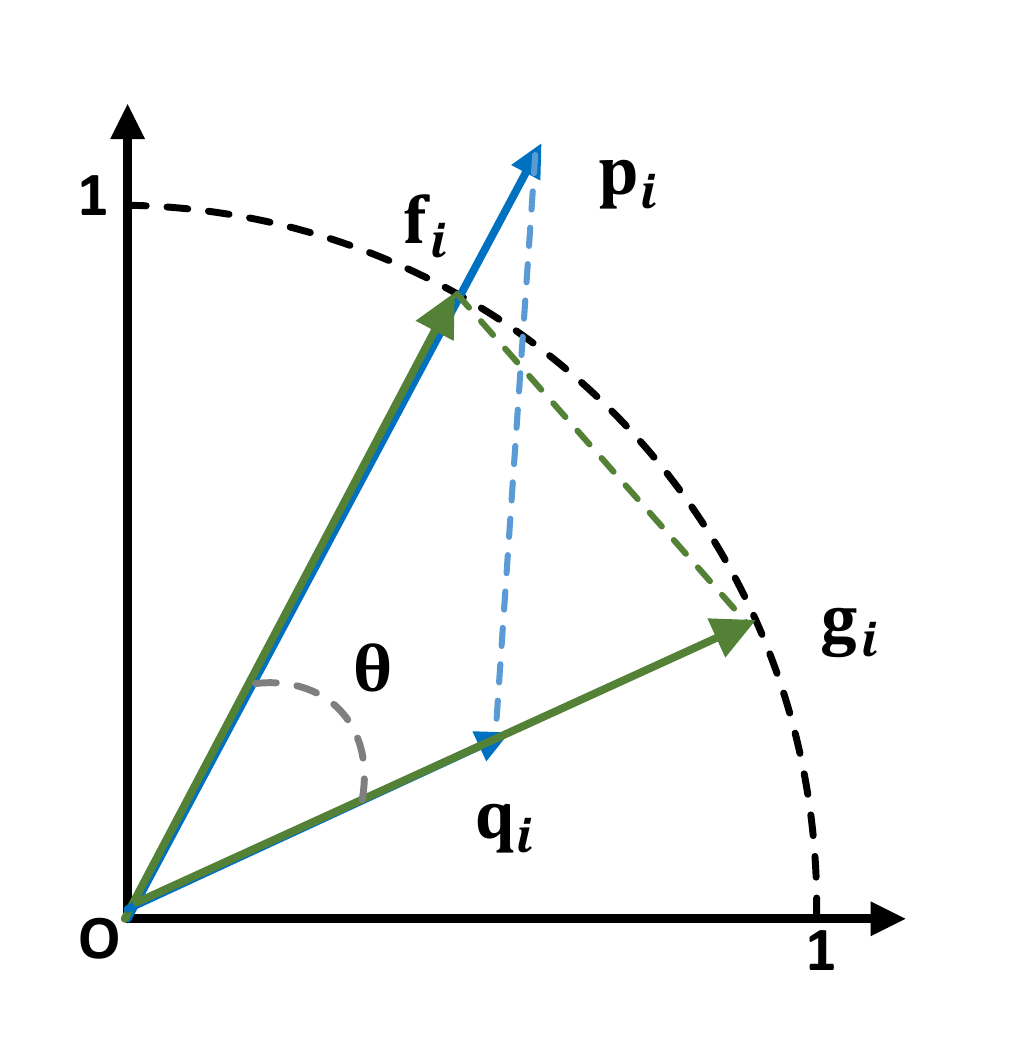}}
	\label{1a}\hfill
	\subfigure[Adjusted cosine distance.]{
		\includegraphics[width=0.36\linewidth]{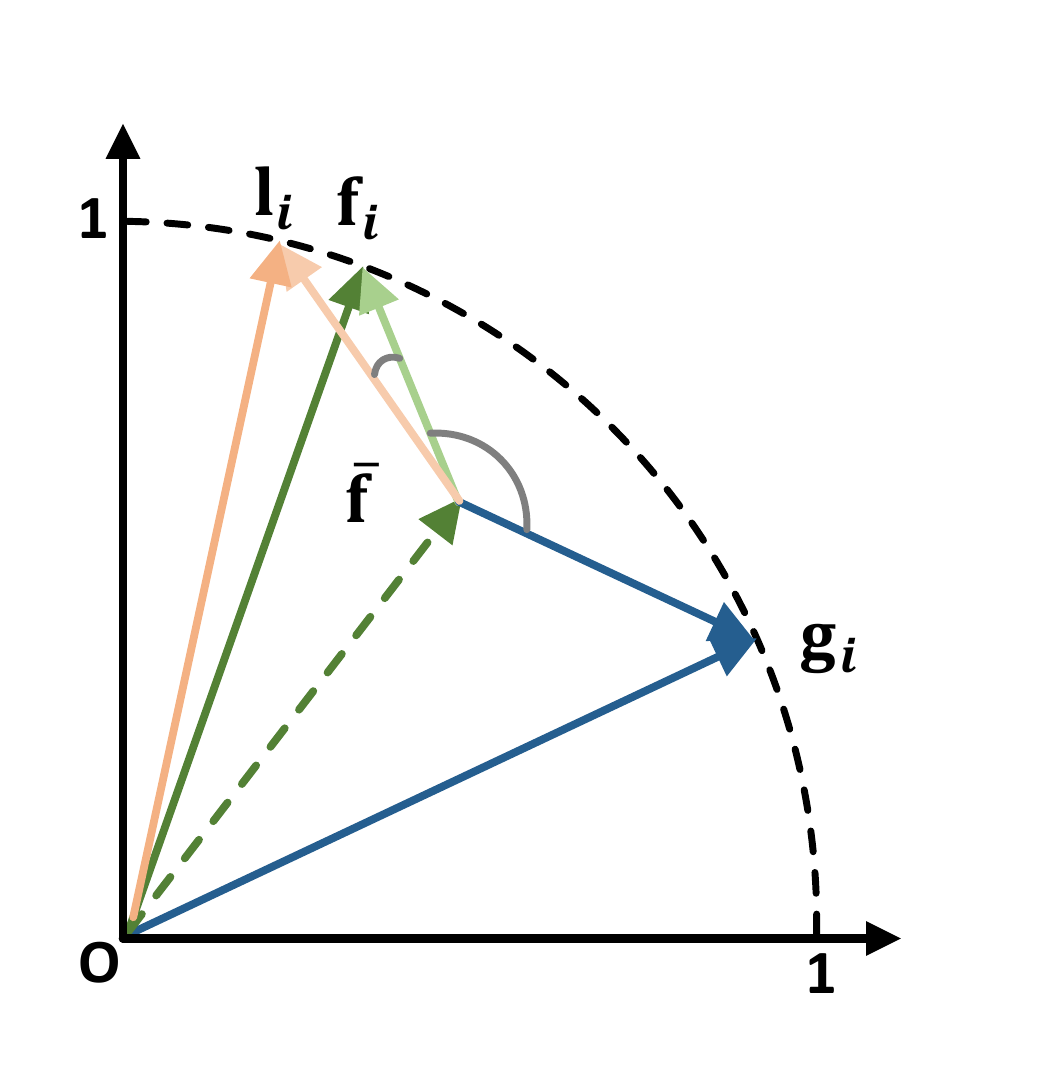}}
	\label{1b}\\
	\caption{Illustration of the cosine distance and its adjusted version.}
	\label{fig:distancemeasurement} 
\end{figure}

After obtaining signal profiles of all robots, ScatterID proceeds to measure distances between their profiles for subsequent identity inference.

\textbf{Distance Matrix Construction.} Consider that there are $ N $ robots who communicate with the ScatterID robot, and a set of their signal profiles can be denoted as $\left\lbrace  \mathbf{F}_1, \mathbf{F}_2, \cdots, \mathbf{F}_N \right\rbrace $. Based on these profiles, we measure the distance between every two profiles and thus obtain a distance matrix $ \mathbf{D} $ as
\begin{equation}
\mathbf{D}=\left[ \begin{array}{cccc}
0      & \mathbf{d}_{12}      & \cdots & \mathbf{d}_{1N}      \\
\mathbf{d}_{21}     & 0      & \cdots & \mathbf{d}_{2N}     \\
\vdots & \vdots & \ddots & \vdots \\
\mathbf{d}_{N1}      & \mathbf{d}_{N2}      & \cdots & 0      \\
\end{array} 
\right ],
\end{equation}
where $ \mathbf{d}_{ij} $ is the distance measurement with respect to two profiles $ \mathbf{F}_i $ and $ \mathbf{F}_j $. Since a signal profile is comprised of $ L $ normalized backscatter signatures, ScatterID calculates the distances between rows of two signal profiles and thereafter outputs a $ L $-dimensional distance vector as 
\begin{align}\label{distance vector}
	\mathbf{d}_{ij}=(d_1,d_2,\cdots, d_L),
\end{align}   
where $ d_l=dist( \mathbf{f}_{il},  \mathbf{f}_{jl}) $, the distance between $l^{\text{th}}$ rows of $ \mathbf{F}_i $ and $ \mathbf{F}_j $. Compared to a scalar distance, the distance vector $ \mathbf{d}_{ij} $ contains the multipath signature differences between two moving robots within a certain period, and thus provides finer-grained information about their long-term similarity. 

\textbf{Adjusted Cosine Distance.} Since a distance vector $ \mathbf{d} $ represents the differences between two signal profiles and is the processing unit for subsequent similarity estimation, the distance metric $ dist(\cdot, \cdot) $ in Eq.~\eqref{distance vector} should be carefully determined. As shown in Fig.~\ref{fig:distancemeasurement}~(a), after the signature normalization, two signature vectors $ \mathbf{p}_i $ and $ \mathbf{q}_i $ are both transformed into unit vectors $ \mathbf{f}_i $ and $ \mathbf{g}_i $ in the feature space, respectively. Such transformation greatly changes the lengths of two signatures and distorts their euclidean distance, making the euclidean metric ineffective in capturing their similarity information. 

To find an effective distance metric, we observe that after the signature normalization, although the lengths of signatures are distorted, their directions remain unchanged. Specifically, the directions of the normalized signatures $ \mathbf{f}_i $ and $ \mathbf{g}_i $ are the same with those of $ \mathbf{p}_i $ and $ \mathbf{q}_i $. Hence, instead of using the length difference, our system leverages the directional difference between two normalized signatures for distance measurement. Traditionally, the cosine distance is used to measure the directional difference between two vectors. As depicted in Fig.~\ref{fig:distancemeasurement}~(a), the cosine distance between $ \mathbf{f}_i $ and $ \mathbf{g}_i $ can be computed as
\begin{align}\label{cosine distance}
 dist_{cos}(\mathbf{f}_i,\mathbf{g}_i)  = 1-\cos(\mathbf{f}_i,\mathbf{g}_i)
 =1- \cos\theta.
\end{align}
Generally, the cosine distance lies in an interval of $ [0,2] $. However, as normalized signatures $ \mathbf{f}_i $ and $ \mathbf{g}_i $ are nonnegative vectors, their angular difference $ \theta $ falls between 0 and $ \pi/2 $, and thus their cosine distance is limited to an smaller interval of $ [0,1] $, which provides fewer discriminative features to differentiate two different signatures.

To address this issue, we propose an adjusted cosine distance for effective angular distance measurement. Formally, for two signal profiles $ \mathbf{F} =[\mathbf{f}_1; \mathbf{f}_2;\cdots; \mathbf{f}_L] $ and $ \mathbf{G}=[\mathbf{g}_1; \mathbf{g}_2;\cdots; \mathbf{g}_L] $ corresponding to two robots, we define the adjusted cosine distance between $ \mathbf{f}_i $ and $ \mathbf{g}_i $ as
\begin{align} \label{adjusted cosine distance}
	dist_{adcos} (\mathbf{f}_i,\mathbf{g}_i) = 1- \cos (\mathbf{f}_i-\mathbf{\bar{f}},\mathbf{g}_i-\mathbf{\bar{f}}),
\end{align}
where $ \mathbf{\bar{f}}=\frac{\mathbf{f}_1 + \mathbf{f}_2 + \cdots + \mathbf{f}_L}{L} $ is the mean vector of $ \mathbf{F} $. According to Eq.~\eqref{adjusted cosine distance}, the newly defined distance is the cosine distance between two vectors $ \mathbf{f}_i - \mathbf{\bar{f}}$ and $ \mathbf{g}_i- \mathbf{\bar{f}} $. As show in Fig.~\ref{fig:distancemeasurement}~(b), for two fake robots, their normalized signatures, i.e., $ \mathbf{l}_i $ and $ \mathbf{f}_i $, should be very close on the unit circle, and their adjusted cosine distance is still too small. While for two normalized signatures from different robots, i.e., $ \mathbf{g}_i $ and $ \mathbf{f}_i $, the adjusted distance will significantly enlarge their directional difference, and therefore a greater angular distance can be obtained. Hence, the adjusted cosine distance provides more discriminative information than the cosine distance in profile distance measurement. Note that since $ dist_{adcos} (\mathbf{g}_i,\mathbf{f}_i) = 1- \cos (\mathbf{g}_i-\mathbf{\bar{g}},\mathbf{f}_i-\mathbf{\bar{g}})$, we have $ dist_{adcos} (\mathbf{f}_i,\mathbf{g}_i) \ne dist_{adcos} (\mathbf{g}_i,\mathbf{f}_i) $ and $ \mathbf{d}_{ij} \ne \mathbf{d}_{ji}$. 

\subsection{Fake Robot Detection}

Based on the measured distance matrix $ \mathbf{D} $, we estimate the similarity between every two robots and further identify fake robots spoofed by Sybil attackers.

\textbf{Similarity Estimation.} Given a distance vector $ \mathbf{d}_{ij} $ between signal profiles of two robots $ i $ and $ j $, we would like to have a similarity probability $ s_{ij} $ as
\begin{align} \label{similarity probability}
	s_{ij}= \mathcal{P}(i=j|\mathbf{d}_{ij}).
\end{align}
Here, $ s_{ij} $ approaches 0 for two different robots and 1 for two fake robots generated by the same attacker, respectively. We achieve this goal by using a machine learning algorithm to fit the conditional distribution $ \mathcal{P}(\cdot|\cdot) $ in Eq.~\eqref{similarity probability}. In particular, the selected algorithm needs to be computationally efficient, because the number of distance vectors in the distance matrix $ \mathbf{D} $ increases as a power function of the number of robots $ N $. Besides, it should be robust to outliers incurred by robotic movements and environmental dynamics.

\begin{figure}
	\centering
	\includegraphics[width=0.88\linewidth]{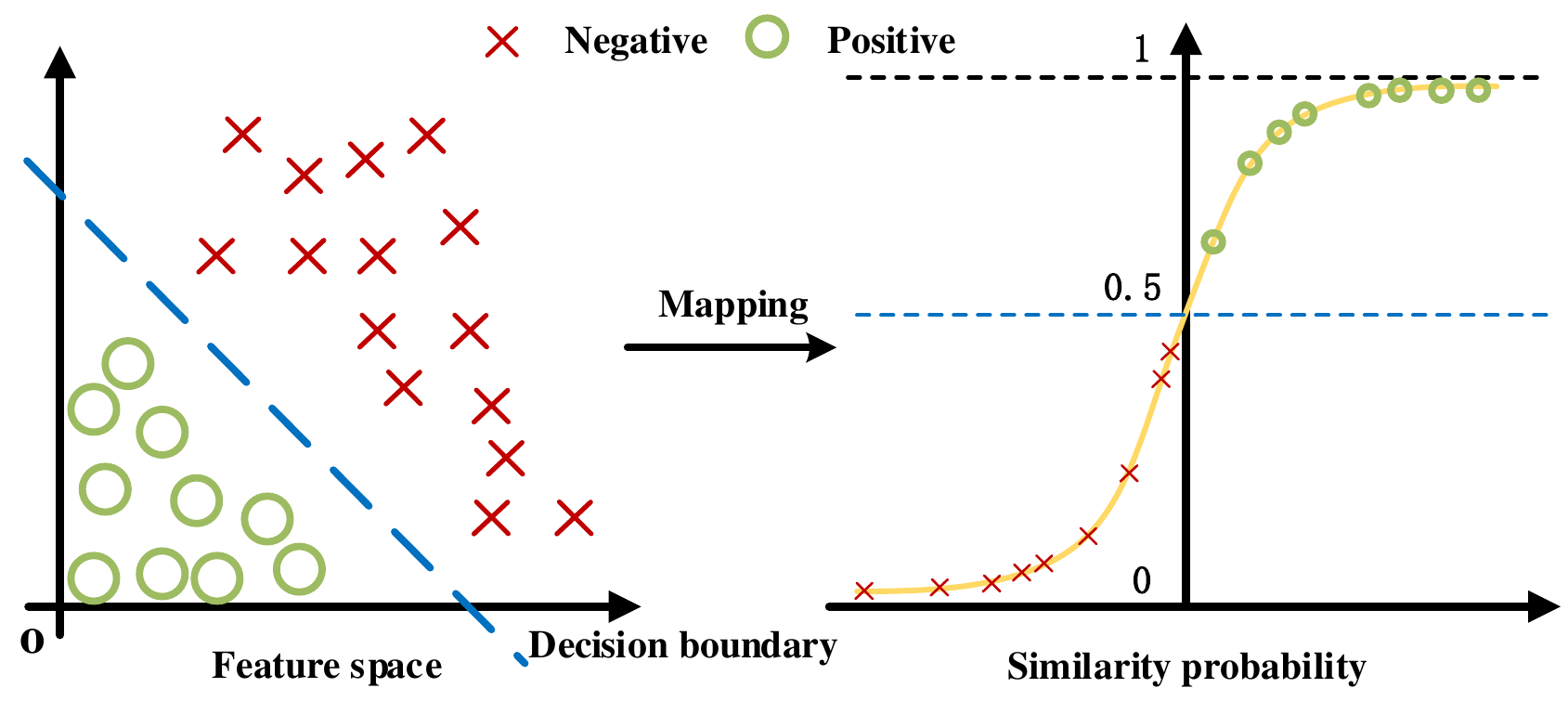}
	\caption{Similarity estimation. The LR model is exploited to find the decision boundary and further estimate the similarity probability between two robots.}
	\label{fig:lr}
\end{figure}

To satisfy these requirements, we leverage a logistic regression (LR) model to estimate the similarity probability $ s_{ij} $ between robots $ i $ and $ j $ based on $ \mathbf{d}_{ij} $ as
\begin{align}
	s_{ij}=g(\mathbf{w}\cdot\mathbf{d}_{ij}+b),
\end{align}
where $ \mathbf{w} $ and $ b $ represent the weight and bias parameters, which together determine a decision boundary in the feature space. Moreover, $ g(\cdot) $ is the sigmoid function, which is given by
\begin{align}
	g(z)=\frac{1}{1+e^{-z}}.
\end{align}
LR is a generalized linear model, mapping the weighted result $ z=\mathbf{w}\cdot\mathbf{d}_{ij} + b $ to a score between 0 and 1 through the sigmoid function, which is suitable in estimating similarity probabilities between two robots. Since the signal profiles of fake robots generated by the same attacker are similar with each other, their distance vectors will almost have small values and locate below the decision boundary in the feature space as illustrated in Fig.~\ref{fig:lr}. Thus, a distance vector with many small values implies a high similarity probability and vice versa. Besides, LR is a highly computationally-efficient algorithm, in which only linear parameters  $ \mathbf{w} $ and $ b $ are needed to learn in the training phase. Moreover, it also takes advantages of the sigmoid function to overcome the unevenness of data distribution and suppress the impact of outliers, which significantly boosts its robustness to robotic movements and environmental dynamics.

In the training phase, we adopt the maximum weighted likelihood estimation (MWLE) to inference the parameters $ \mathbf{w} $ and $ b $. Specifically, in MWLE, the goal is to maximize its weighted likelihood function $ \mathcal{L}(\mathbf{w}, b) $ with respect to $ \mathbf{w} $ and $ b $, which is defined as
\begin{align}
	\mathcal{L}(\mathbf{w}, b)=\prod_{n=1}^{N} v_n  [g(z_n)]^{y_n}  [1-g(z_n)]^{1-y_n},
\end{align}
where $ z_n=\mathbf{w}\cdot\mathbf{d}_n+b $. Therein, $ N $ is the number of training samples, $ \mathbf{d}_n $ a distance sample, $ y_n \in \left\lbrace 0, 1\right\rbrace  $ the sample label, and $ v_n $ the sample weight. In our application, $ y_n = 1 $ flags a distance vector between two fake robots and $ y_n=0 $ a distance vector between a different pair. Moreover, to alleviate the class imbalance problem, we set the sample weight $ v_n $ to be inversely proportional to the number of same-label samples.

\textbf{Robot Identification.} After inputting a distance vector into the LR model, we obtain a similarity probability between a pair of robots. The closer the probability to one, the more likely the robotic pair comes from the same moving robot. Thus, after inputting all elements in the distance matrix $ \mathbf{D} $, we can obtain a similarity matrix $ \mathbf{S} $ for all robots as
\begin{equation}
	\mathbf{S}=\left[ \begin{array}{cccc}
0      & s_{12}      & \cdots & s_{1N}      \\
s_{21}     & 0      & \cdots & s_{2N}     \\
\vdots & \vdots & \ddots & \vdots \\
s_{N1}      & s_{N2}      & \cdots & 0      \\
	\end{array} 
	\right ].
\end{equation}
Recall that for two robots, their distance vectors $ \mathbf{d}_{ij} $ and $ \mathbf{d}_{ji} $ are different, thus the corresponding similarity probabilities $ s_{ij} $ and $ s_{ji} $ are not same, which indicates that $ \mathbf{S} $ is an asymmetric matrix. 

Based on the similarity matrix $ \mathbf{S} $, we determine that two robots $ i $ and $ j $ are spoofed by the same Sybil attacker when
\begin{align}
	 s_{ij} \geq \sigma  \quad  \text{and} \quad  s_{ji}\geq \sigma, 
\end{align}
where $ \sigma \in \left(  0,1 \right)  $ is the similarity threshold for determination. Considering that $ s_{ij} $ and $ s_{ji} $ are two probabilities, we set the threshold $ \sigma $ to be 0.5.

\begin{figure*}
	\hfill
	\begin{minipage}[t]{0.35\linewidth}
		\includegraphics[width=0.99\linewidth]{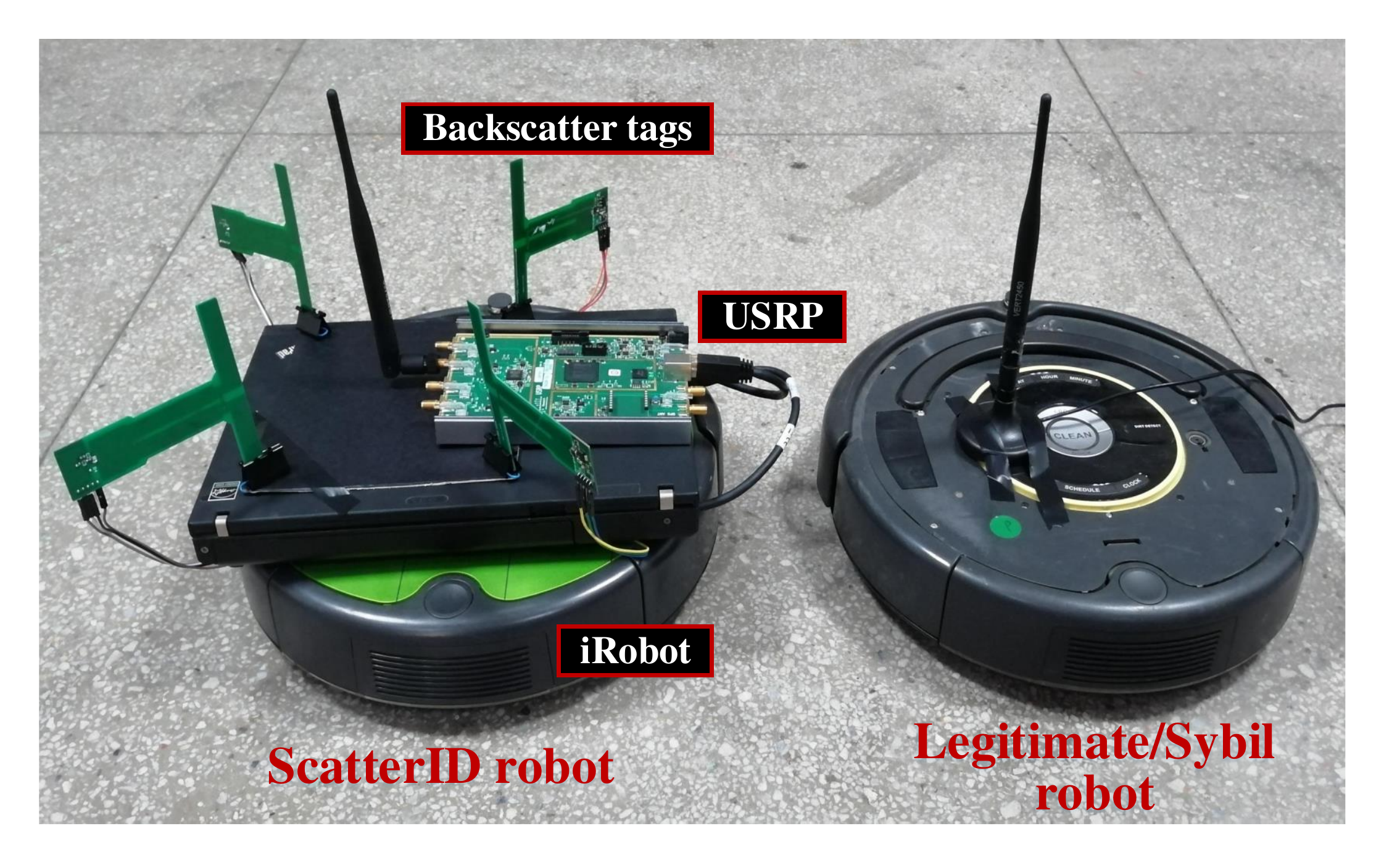}
		\caption{Experimental platform. It involves USRP nodes, backscatter tags and iRobot Create robots.}
		\label{fig:platform}
	\end{minipage}
	\hfill
	\begin{minipage}[t]{0.325\linewidth}
		\centering
		\includegraphics[width=0.99\linewidth]{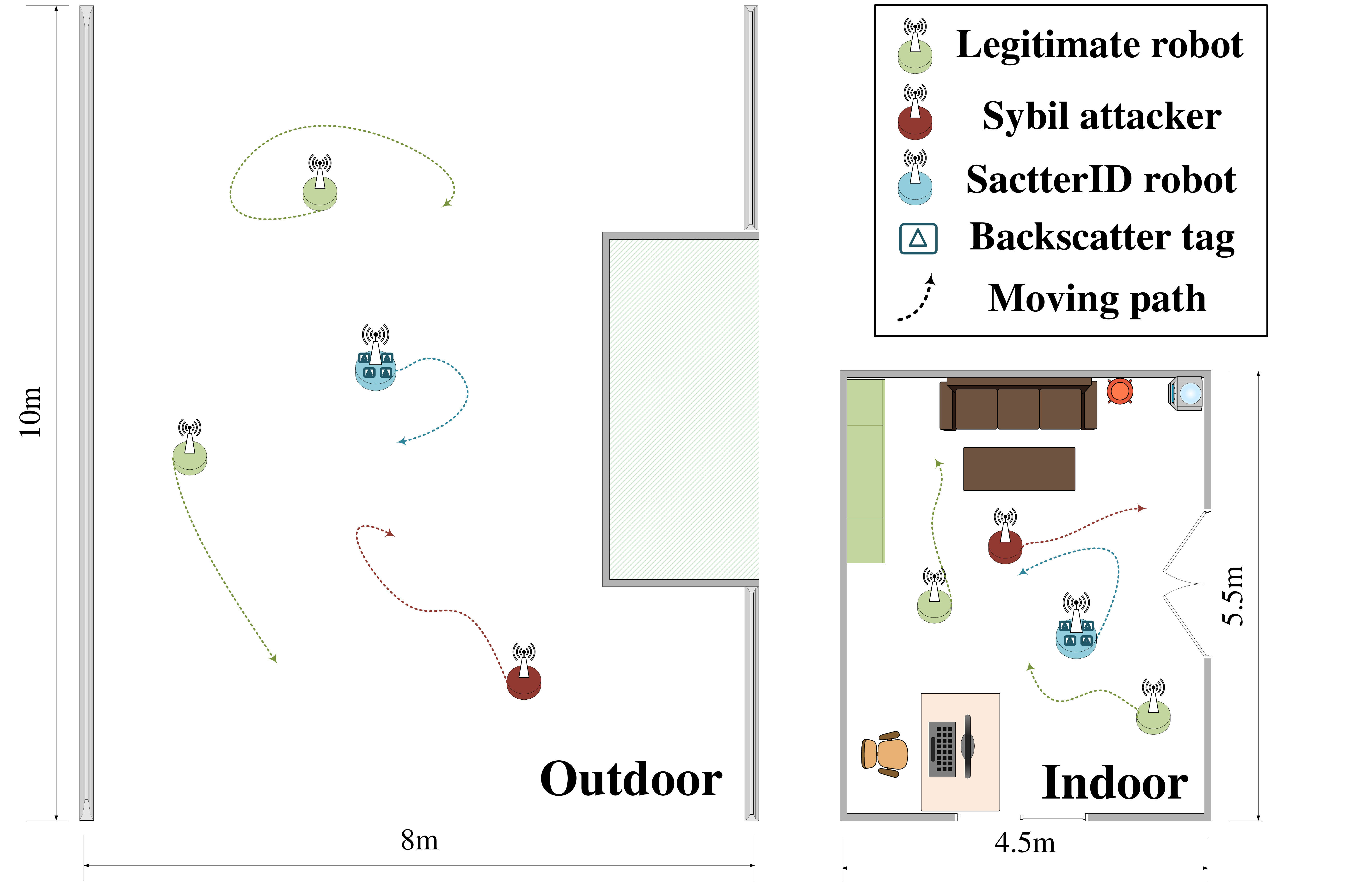}
		\caption{Experimental environments. We run our system in indoor and outdoor environments.}
		\label{fig:environmentsetting}
	\end{minipage}
	\hfill
	\begin{minipage}[t]{0.31\linewidth}
		\centering
		\includegraphics[width=0.99\textwidth]{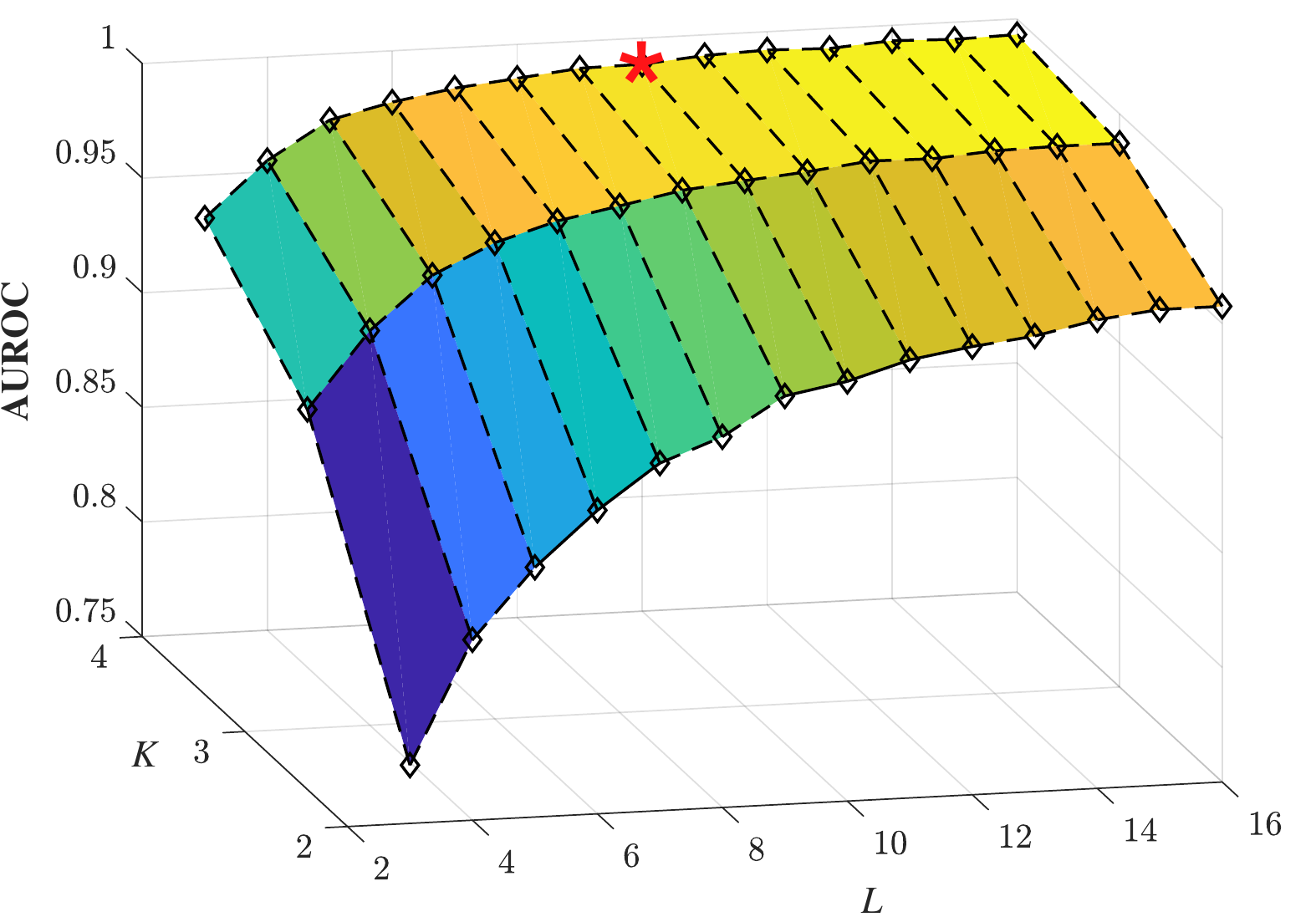}
		\caption{AUROCs under different profile sizes. The red mark corresponds to the selected size.}\label{fig:tagnumsfig}
	\end{minipage}
\end{figure*}

\section{Implementation and Evaluation} \label{sec:experiment}

\subsection{Implementation}
As shown in Fig.~\ref{fig:platform}, we implement our system using iRobot Create robots, backscatter tags, and GNURadio/USRP B210 nodes. We prototype our backscatter tags using off-the-shelf circuit components based on the design in~\cite{liu2013ambient}. Tags are controlled by an Altera STEP-MAX10 FPGA to reflect signals with a frequency shift of 20~MHz to avoid interference and transmit data with a bitrate of 4 Kbps. We build a ScatterID robot on an iRobot Create platform equipped with multiple backscatter tags and a USRP node with a single antenna. The USRP node is surrounded by tags at a distance of 12~cm. Other iRobot Create platforms each equipped with one USRP node act as Sybil attackers or legitimate robots. All robots are configured to communicate in the 2.4~GHz ISM band. 

\subsection{Evaluation Methodology}
\textbf{Experimental Setup.} We evaluate the performance of ScatterID in both indoor and outdoor environments, including an office room with a size of 4.5~m $ \times $ 5.5~m and a building rooftop with an area of 8~m $ \times $ 10~m, as shown in Fig.~\ref{fig:environmentsetting}. The office room is a typical multipath-rich environment, which contains some furniture and surrounding walls. The rooftop is a multipath-poor area with a few walls and roof guardrails. In our experiments, we set the ScatterID robot to update multipath signatures every 0.6~s. In addition, to launch power-scaling attacks, a robotic attacker can change its transmission power for different IDs. Moreover, each robot has a different moving path with a speed of 20~cm/s in two settings.

\textbf{Dataset.} In our experiments, we collect a total of four-hour of backscattered signal traces over different days. Based on the collected traces, we extract signal profiles of all robots and measure their distance vectors as described in Section~\ref{sec:system design}. We obtain our dataset that includes more than 66K distance samples. Therein, about one fifth of them are positive and the left are negative. We randomly partition our dataset into 10 subsets and use 10-fold cross-validation for training and testing our LR model. In each subset, samples from indoor and outdoor environments are included. 

\textbf{Evaluation Metrics.} We use the following metrics to evaluate the performance of our system.
\begin{itemize}
	\item \textbf{Accuracy.} It is computed as the ratio of the total number of robots that are correctly recognized to the total number of legitimate and fake robots.
	\item \textbf{True positive rate (TPR).}  It is the ratio of the number of fake robots that are successfully detected to the total number of fake robots. 
	\item \textbf{False positive rate (FPR).} It is the ratio of the number of legitimate robots that are mistakenly recognized to the total number of legitimate robots.
	\item \textbf{Receiver operating characteristic (ROC) curve.} It is a curve in terms of TPRs and FPRs with varying discrimination thresholds in the interval $ \left[0,1 \right]  $.
	\item \textbf{AUROC.} It is defined as the area under ROC curve, falling into $ \left[0.5,1 \right]  $. The higher the value is, the better performance is achieved.
\end{itemize}

\subsection{Results}

\textbf{Profile Size Determination.} As a signal profile is an essential unit for attack detection, the first step of our experiments is to determine its size, which is decided by $ K $ and $ L $, the numbers of backscatter tags and normalized signatures, respectively. Generally, the larger the profile size is, the better performance will be achieved by our system. However, increasing profile size will also incur more computation complexity and time. Thus, a small-size profile that guarantees high performance is desired. In order to find an appropriate profile size, we record different AUROC values when changing $ K $ and $ L $ from 2 to 4 and 16, respectively. As depicted in Fig.~\ref{fig:tagnumsfig}, we observe that AUROC grows from the lowest value 0.775 to the highest 0.993 as $ K $ and $ L $ both increase. Moreover, for the number of tags, four tags ensure that all AUROC values are more than 0.9. For the number of signatures, AUROC grows quickly at the beginning and has marginal growth when $ L $ exceeds 10. Therefore, we take ten successive normalized signatures from four backscatter tags as a signal profile.

\begin{figure*}
	\hfill
	\begin{minipage}[t]{0.26\linewidth}
		\centering
		\includegraphics[width=0.99\textwidth]{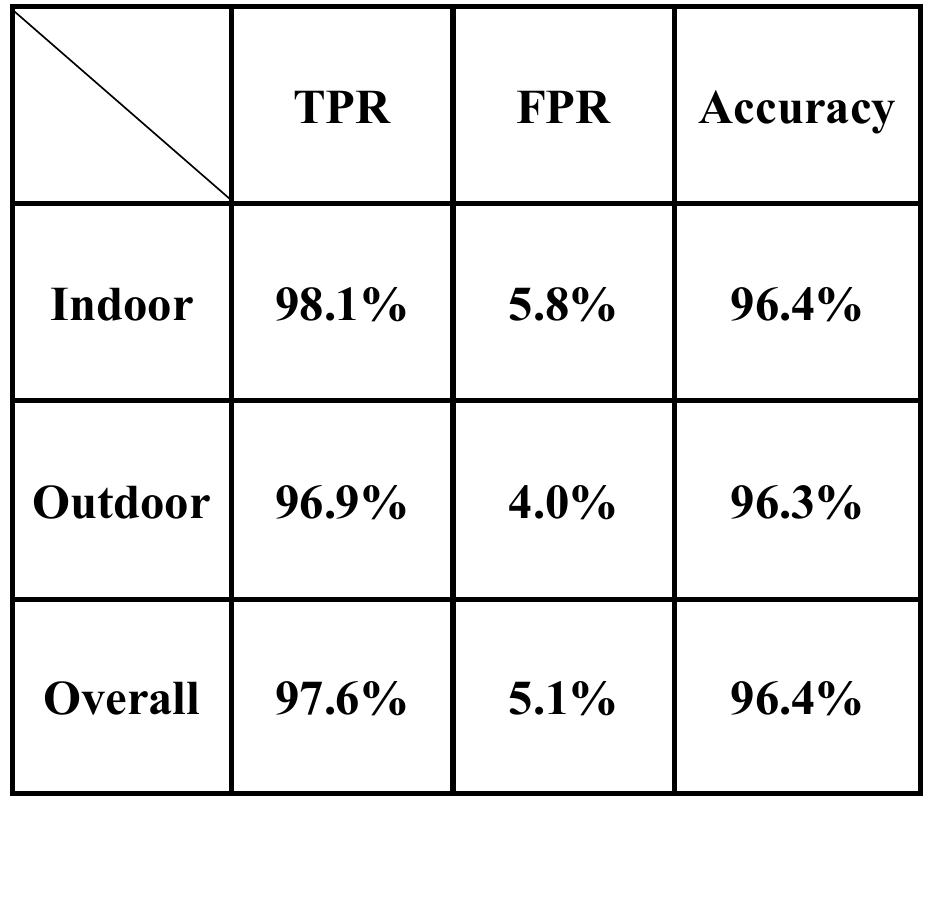}
		\caption{System performance in indoor and outdoor environments.}\label{fig:overallperformance}
	\end{minipage}
	\hfill
	\begin{minipage}[t]{0.29\linewidth}
		\centering
		\includegraphics[width=0.90\textwidth]{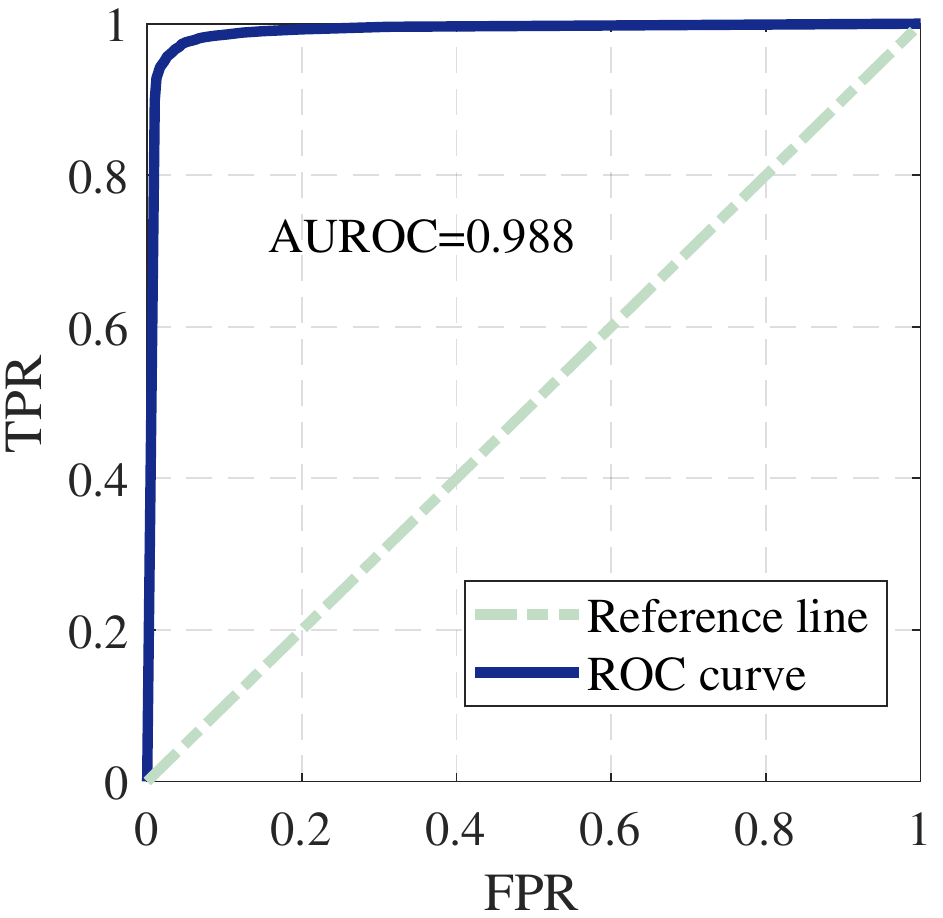}
		\caption{The ROC curve of our system. The corresponding AUROC reaches to 0.988. }\label{fig:roccurve}
	\end{minipage}
	\hfill
	\begin{minipage}[t]{0.43\linewidth}
		\centering
		\includegraphics[width=0.99\textwidth]{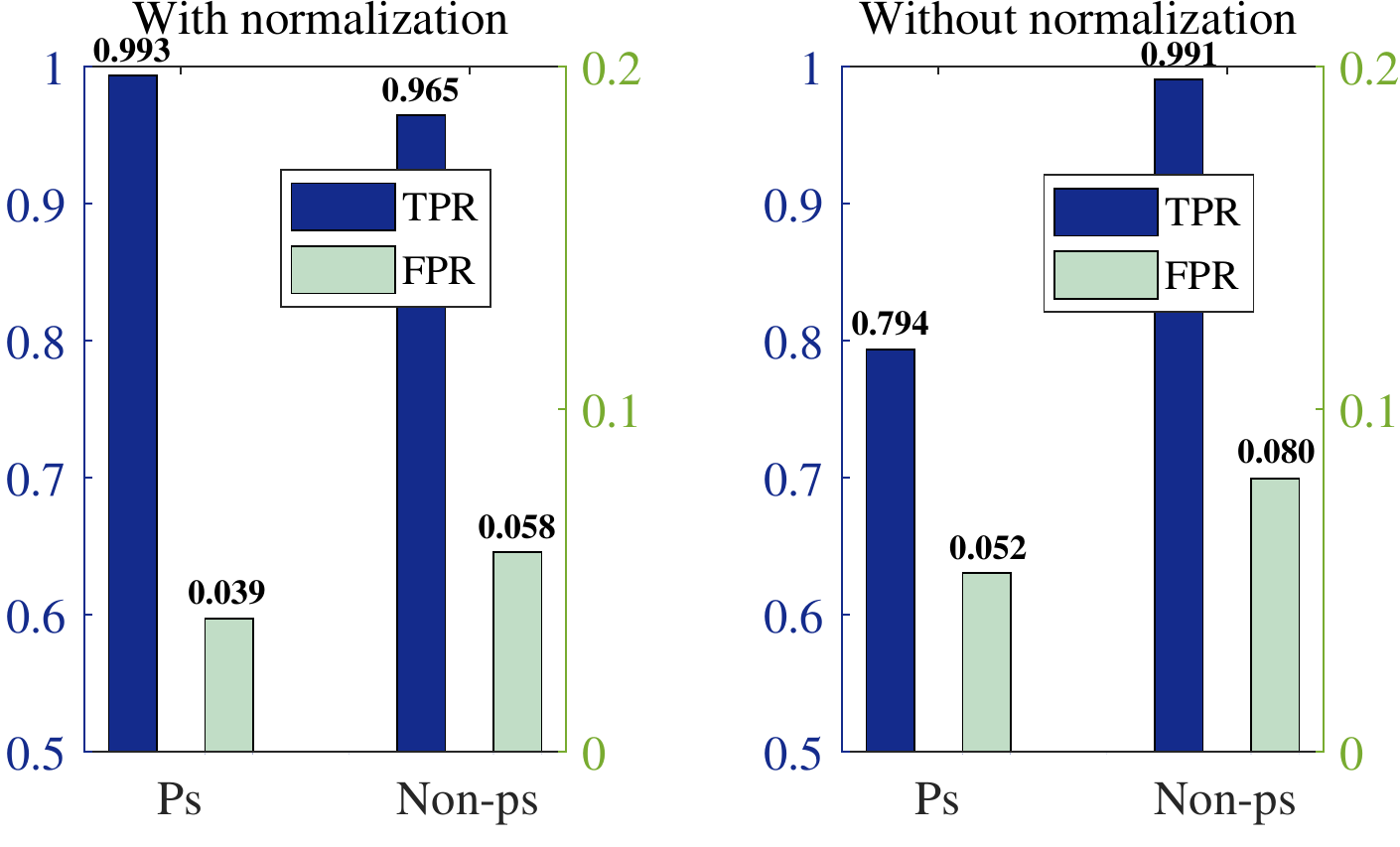}
		\caption{TPR and FPR with and without signature normalization. Ps indicates the power-scaling scenario.}\label{fig:signaturenormalization}
	\end{minipage}
\end{figure*}

\textbf{Overall Performance.} Then, we report the overall performance of ScatterID in Fig.~\ref{fig:overallperformance} and Fig.~\ref{fig:roccurve}. We first summarize experimental results in indoor and outdoor environments with respect to TPR, FPR and accuracy. Compared with the outdoor environment, the indoor environment has much richer multipath due to scattering and reflections from surrounding furniture items and walls. As shown in Fig.~\ref{fig:overallperformance}, such multipath effects make our system obtain higher TPR and FPR in the indoor environment. However, when it comes to accuracy, our system achieves consistent performance in two different settings. To sum up, our system has a detection accuracy of 96.4\%. In particular, it can successfully mitigate 97.6\% of Sybil attacks and correctly recognize 94.9\% of legitimate traffic at the same time. Next, we present the ROC curve of our system. As shown in Fig.~\ref{fig:roccurve}, the ROC curve closely follows the left-hand border and the top border of the ROC space and produces a high AUROC of 0.988, which is close to the ideal case. The above results show the effectiveness and robustness of our system in discriminating between legitimate and fake robots.

\textbf{Impact of Signature Normalization.} Next, we show that our system is resilient to power-scaling attacks, through which an attacker can easily imitate different robots. Fig.~\ref{fig:signaturenormalization} illustrates the impact of signature normalization when attackers perform Sybil attacks with and without varying transmission power. As shown in Fig.~\ref{fig:signaturenormalization}, we observe that in the power-scaling scenario, the signature normalization can significantly improve TPR by approximately 20\% while reducing FPR by about 2\%. This is because in each transmission, backscattered signal strengths are proportional to the transmission power, and thus the varying coefficient can be eliminated via normalization operations. However, in the non-power-scaling scenario, the normalization operation has no significant impact on both TPR and FPR. Moreover, with signature normalization, our system achieves comparable performance in both power-scaling and non-power-scaling scenarios. Based on the above observations, we conclude that our system is robust to Sybil attackers with power-scaling ability.

\begin{figure*}
	\hfill
	\begin{minipage}[t]{0.35\linewidth}
		\centering
		\includegraphics[width=0.98\linewidth]{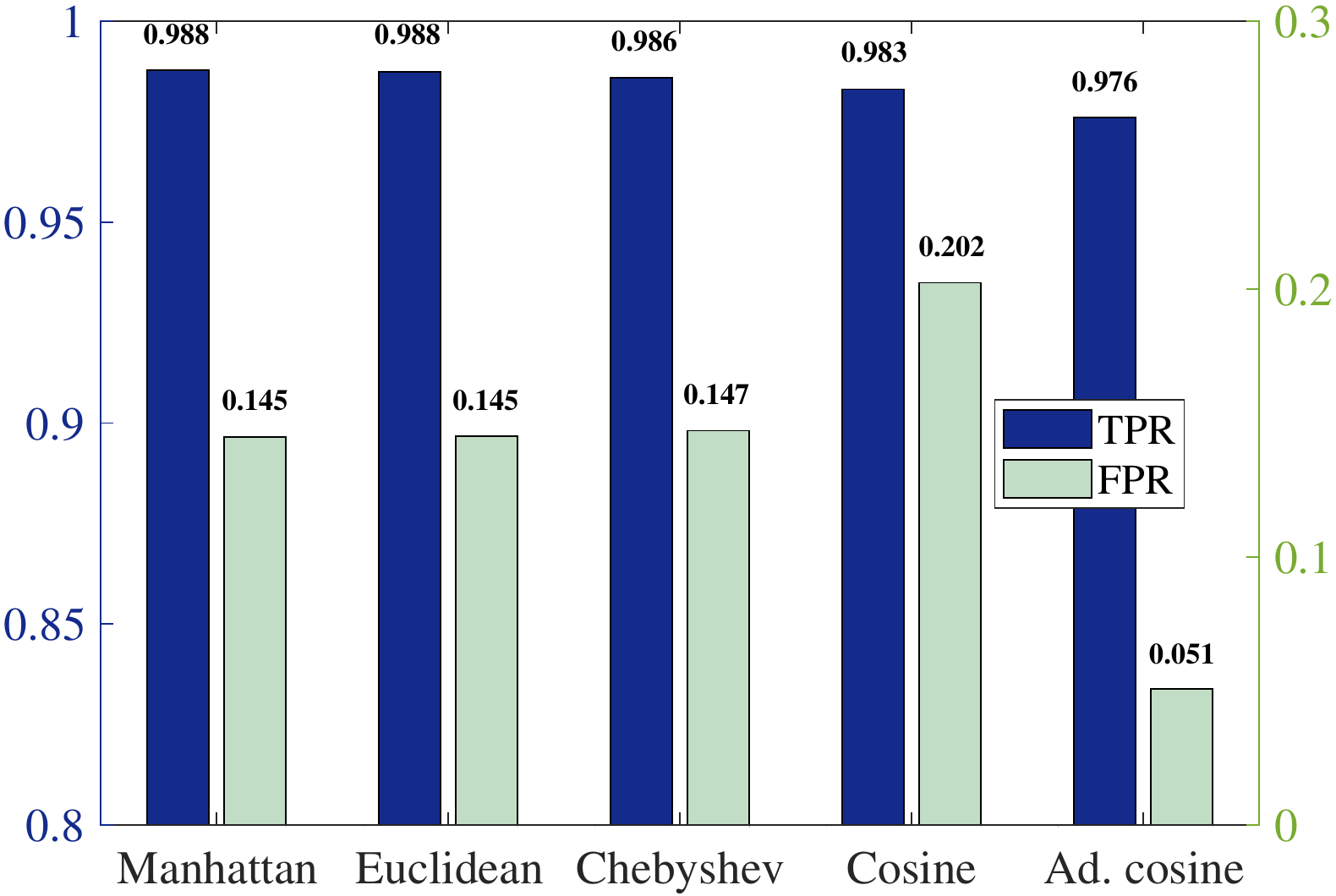}
		\caption{TPR and FPR under different distance metrics.}
		\label{fig:distancefunctions}
	\end{minipage}
	\hfill
	\begin{minipage}[t]{0.315\linewidth}
		\centering
		\includegraphics[width=0.97\linewidth]{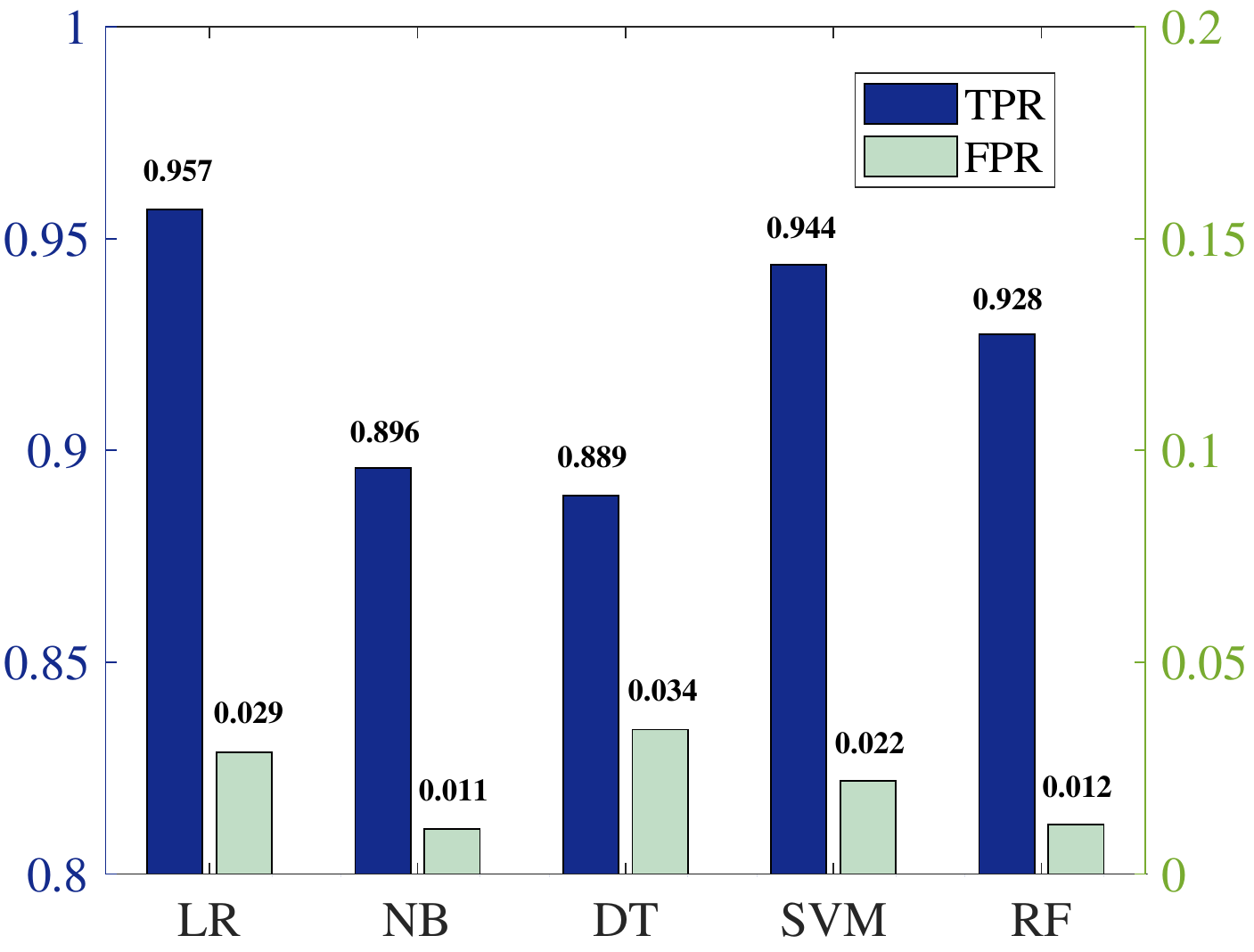}
		\caption{Binary classification results of different machine learning models.}
		\label{fig:fptprates}
	\end{minipage}
	\hfill
	\begin{minipage}[t]{0.315\linewidth}
		\centering
		\includegraphics[width=0.97\linewidth]{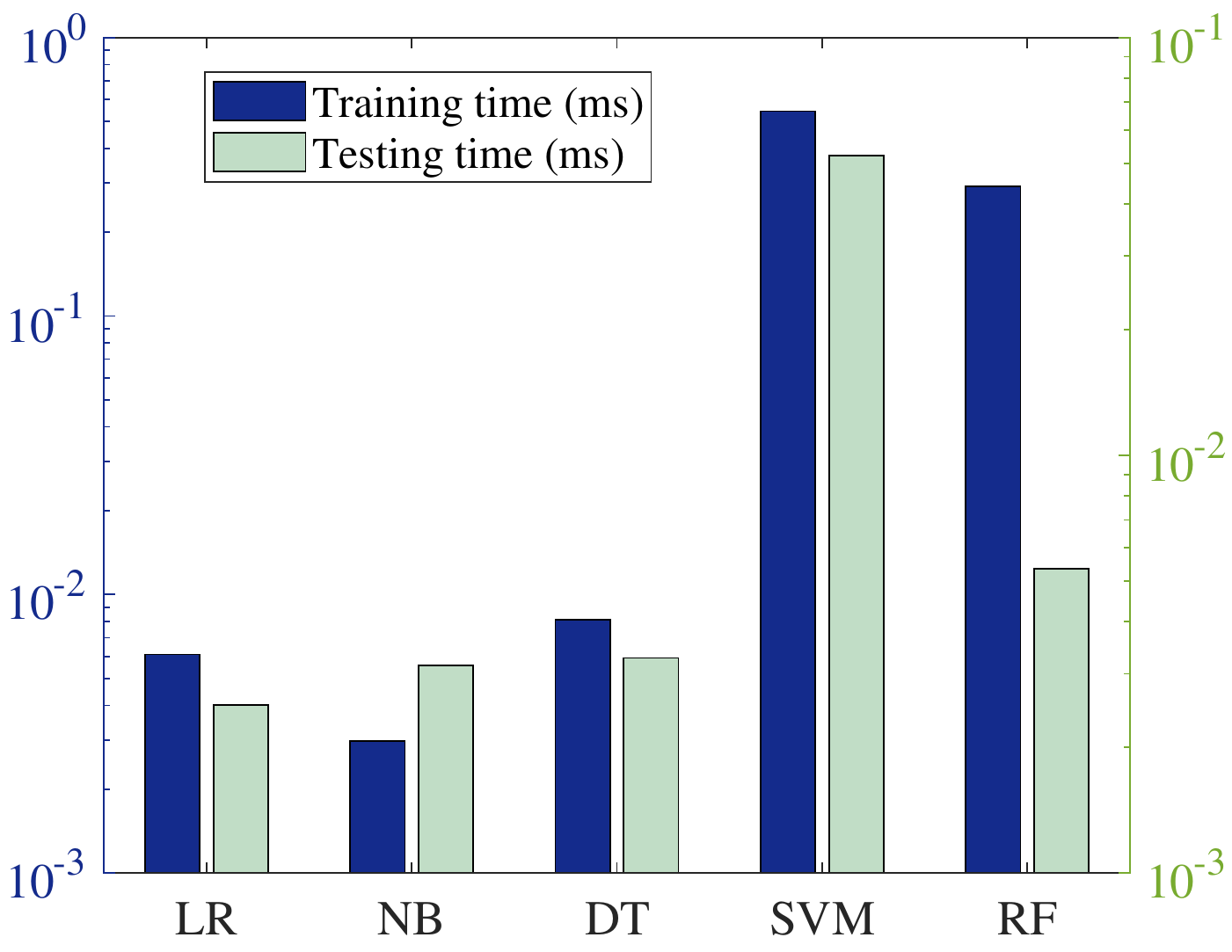}
		\caption{Training and testing time of different machine learning models for one sample.}
		\label{fig:trainingtestingtime}
	\end{minipage}
\end{figure*}

\textbf{Impact of Distance Metrics.} We further check the effectiveness of the adjusted Cosine distance in profile distance measurement. To this end, we compare the proposed distance with traditional metrics, i.e., \textit{Manhattan}, \textit{Euclidean}, \textit{Chebyshev} and \textit{Cosine} distances, and plots the results in Fig.~\ref{fig:distancefunctions}. We can observe that Manhattan, Euclidean and Chebyshev distances show similar performance with a TPR of about 98\% and a FPR of more than 14\% each. Despite the relatively high TPRs, their FPRs are very high too, which makes them inapplicable in reliably recognizing legitimate robots. Moreover, the FPR corresponding to cosine distance is more than 20\%. The reason is that the cosine distance is limited to $ \left[0,1 \right]  $ in our application and thus provides less discriminative information between two different profiles. However, the adjusted cosine distance enables our system to achieve not only a relatively high TPR of 97.6\% but also a low FP rate of 5.1\%. This is due to the fact that the adjusted cosine distance reliably enlarges the difference between a pair of different profiles, while it has nearly no impact on similar profiles. Thus, the adjusted cosine distance is an effective metric in profile distance measurement.

\textbf{Impact of Machine Learning Algorithms.} Finally, we present the benefits of adopting the LR model in our system. We compare LR with other learning models by considering binary classification performance as well as computational overhead for each distance sample, and report the results in Fig.~\ref{fig:fptprates} and Fig.~\ref{fig:trainingtestingtime}, respectively. In the experiment, the baseline models are comprised of naive Bayes (NB), decision tree (DT), support vector machine (SVM) and random forest (RF). As shown in Fig.~\ref{fig:fptprates}, all models exhibit negligible differences in FPRs and all of them are lower than 0.04. However, LR, SVM and RF have high TPRs of more than 0.92, and the TPRs of the other two models are below 0.9. When it comes to computational overhead, as depicted in Fig.~\ref{fig:trainingtestingtime}, LR, NB and DT are computational-efficient and have the same level of training and testing time, less than 0.01~ms in both training and testing phases. However, SVM and RF are time-consuming models, which incur an order of magnitude higher computation time, especially in the training phase. By taking above two factors into consideration, we conclude that LR model is most suitable in our application.

\section{Related Work} \label{sec:related work}
\textbf{Multi-Robot Networks.} Recent years have witnessed the proliferation of multi-robot networks in many emerging applications such as surveillance~\cite{rybski2002performance}, exploration~\cite{corah2019distributed} and coverage~\cite{gil2015adaptive,ghaffarkhah2014dynamic}. In these applications, one of the key research issues is how to effectively cooperate among robots in order to achieve a high quality of overall performance. However, effective cooperation is based on the assumption that all claimed identities are accurate and trustful, which can be easily undermined by Sybil attacks.

\textbf{Backscatter Communications.} Nowadays, backscatter has become one of the most energy-efficient communication primitives. In~\cite{liu2013ambient}, ambient backscatter is originally proposed to facilitate two simple and batteryless tags to communicate with each other by reflecting ambient radio signals from surrounding TV towers. After that, a variety of novel techniques, such as WiFi backscatter~\cite{kellogg2015wi-fi}, FS backscatter~\cite{zhang2016enabling}, FM backscatter~\cite{wang2017fm} and LoRa backscatter~\cite{talla2017lora}, are successively developed to achieve high energy efficiency. Beyond their benign uses, backscatter has increasingly been used in other applications such as on-body device authentication~\cite{luo2018authenticating} and object tracking~\cite{luo20193d}. In~\cite{luo2018}, backscatter tags are used to shield static Internet of Things (IoT) networks from spoofing attacks, where attackers try to masquerade other legitimate nodes. Differing from the prior work~\cite{luo2018}, we consider Sybil attacks in dynamic robotic networks and provide effective resilience to power-scaling attacks.

\textbf{Sybil Attack Mitigation.} Sybil attacks have been widely considered in multi-node networks~\cite{newsome2004sybil,faria2006detecting}. The past solutions on Sybil attack mitigation are mainly falling into two categories. (i) Cryptographic-based approaches~\cite{chan2003random,ramkumar2005an} assume prior trust among network nodes and require computationally expensive PSK management schemes. These requirements, however, cannot be satisfied in ad hoc and miniaturized robotic platforms. (ii) Non-cryptographic approaches~\cite{demirbas2006an,xiao2009channel-based,liu2015the,faria2006detecting,xiong2013securearray,gil2017guaranteeing} use wireless PHY information to discriminate between real and fake nodes without the assumption of mutual trust and excessive computational overhead. However, these approaches passively observe PHY features using bulky multiple antennas and thus do not suit to miniaturized robots with limited payload and hardware capabilities. To transcend these limitations, our work adopts featherlight and batteryless backscatter tags to actively capture fine-grained multipath features and constructs sensitive signal profiles for Sybil attack detection.

\section{Conclusion}\label{sec:conclusion}
This paper presents ScatterID, a lightweight system against Sybil attacks for networked and miniaturized robots in many cooperative tasks. Instead of passively measuring PHY features using bulky multiple antennas, ScatterID utilizes featherlight backscatter tags to actively manipulate multipath propagation and creates salient multipath features obtainable to single-antenna robots. These features are used to identify the spatial uniqueness of each moving robot, even under power-scaling dynamic attacks. We implement our system on commercial off-the-shelf robotic platforms and extensively evaluate it in typical indoor and outdoor environments. The experimental results show that ScatterID achieves a high AUROC of 0.988 and an overall accuracy of 96.4\% for identity verification. In addition, it can successfully detect 97.6\% of fake robots while mistakenly rejecting just 5.1\% of legitimate ones.

\bibliographystyle{IEEEtran}
\bibliography{IEEEabrv,./Sybildetection}

\end{document}